\documentclass{article}

\usepackage{arxiv}

\usepackage[utf8]{inputenc} 
\usepackage[T1]{fontenc}    
\usepackage{hyperref}       
\usepackage{url}            
\usepackage{booktabs}       
\usepackage{amsfonts}       
\usepackage{nicefrac}       
\usepackage{microtype}      
\usepackage{lipsum}

\usepackage{graphicx}
\usepackage{tabularx} 
\usepackage{multirow}

\usepackage{amsmath}

\title{SoK: Privacy-Enhancing Technologies in Artificial Intelligence}

\author{
  Nouha Oualha\\
  Université Paris-Saclay, CEA, List, F-91120, Palaiseau, France\\
  \texttt{nouha.oualha@cea.fr} \\
}

\begin{document}

\maketitle

\begin{abstract}
As artificial intelligence (AI) continues to permeate various sectors, safeguarding personal and sensitive data has become increasingly crucial. To address these concerns, privacy-enhancing technologies (PETs) have emerged as a suite of digital tools that enable data collection and processing while preserving privacy. This paper explores the current landscape of data privacy in the context of AI, reviews the integration of PETs within AI systems, and assesses both their achievements and the challenges that remain.
\end{abstract}


\section{Introduction}
Artificial intelligence (AI) is gaining momentum across several sectors, including manufacturing, transportation, finance, and medicine. In today’s data-driven world, the development and deployment of AI systems often require the collection and processing of vast amounts of personal and sensitive data, raising significant concerns regarding privacy and security.

The concept of privacy by design \cite{Cybersecurity.2023} emphasizes the integration of privacy and data confidentiality into the core architecture of AI systems. This is often achieved through the development and adoption of privacy-enhancing technologies (PETs), which aim to ensure data protection and confidentiality. In this context, privacy-preserving machine learning (PPML) has emerged as a promising field that enhances the privacy and security of sensitive data while enabling effective machine learning (ML) applications. PPML leverages various PETs designed to maintain privacy without compromising model performance.

PETs encompass a broad range of approaches, tools, and concepts that aim to preserve users' privacy and protect the confidentiality—and, in some cases, the integrity and availability—of personal and sensitive data. Integrating PETs into AI systems enables organizations to mitigate the risks associated with unauthorized access, data breaches, and the misuse of sensitive information.

This survey consolidates existing knowledge on privacy and data confidentiality within the AI domain, with a specific focus on ML as a sub-field of AI. It examines the role of PETs in AI systems and identifies key challenges that remain to be addressed.

While this is not the first state-of-the-art review on security and privacy in AI (e.g., \cite{dAliberti2024}, \cite{DeCristofaro2020}, \cite{ElMestari2024}, \cite{Mireshghallah2020}, \cite{PulidoGaytan2021}), this survey provides a more comprehensive overview of the various privacy-preserving technologies applicable to AI/ML. Unlike previous work that focuses on specific AI models (e.g., large language models in \cite{Dong2024}, [8]) or particular privacy concerns (e.g., the right to be forgotten in \cite{Zhang2023}), this paper adopts a broader and more inclusive approach.

\section{Motivation}
PETs play a pivotal role in ensuring compliance with stringent privacy laws and regulations. Violations of these regulations may result in substantial fines. The current regulatory landscape includes the Health Insurance Portability and Accountability Act (HIPAA), the California Consumer Privacy Act (CCPA), Biden’s Executive Order on Safe, Secure, and Trustworthy Artificial Intelligence, and China’s Personal Information Protection Law (PIPL), as well as the European legal framework consisting primarily of the General Data Protection Regulation (GDPR), the Data Governance Act, and the Artificial Intelligence Act \cite{ElMestari2024}.

As reliance on data-driven models expands across various domains, so do concerns regarding user consent. The increasing awareness and demand for privacy-preserving solutions further underscore the need for integrating PETs into AI systems.

PETs provide essential tools for safeguarding privacy without compromising AI functionality. These tools support various privacy guarantees \cite{dAliberti2024}, based on information-theoretic or computational security, including:
\begin{itemize}
	\item Learning from private data without revealing sensitive information.
	\item Performing secure computations and protecting data in untrusted environments.
	\item Safeguarding users' privacy during interaction with ML systems, while also protecting the model's confidentiality from malicious users.
	\item Enabling collaborative parties to share only the minimum required data (i.e., data minimization) and securely exchange AI models without exposing sensitive information.
	\item Allowing data owners to provide informed consent and maintain control over their data.
\end{itemize}

\section{Machine learning overview}

Machine learning (ML), a sub-field of AI, focuses on developing models capable of learning from data to perform specific tasks. Most ML models are parametric, approximating data behaviour through a parametric function $f(x,\theta)$, where $x$ is a feature vector representing input data, and $\theta$ is a parameter vector optimized during training \cite{DeCristofaro2020}.

Training is typically achieved through one of three main paradigms: supervised, unsupervised, or reinforcement learning.
\begin{itemize}
\item Supervised learning involves training models on labeled datasets, enabling the system to learn relationships between inputs and outputs.
\item Unsupervised learning is based on data without labels, allowing the model to identify patterns and structure within the dataset.
\item Reinforcement learning uses trial-and-error learning, where agents make decisions based on rewards or penalties.
\end{itemize}

\begin{figure}[ht!]
	\centering
	\includegraphics[width=0.5\columnwidth]{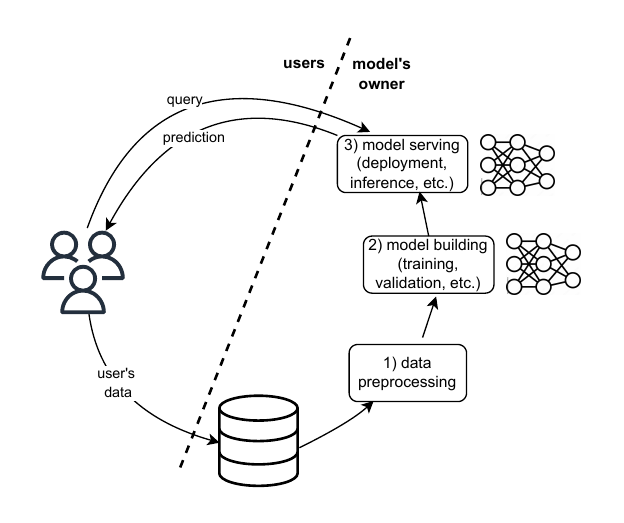}
	\caption{ML phases: 1) data preprocessing, 2) model building, and 3) model serving.}
	\label{fig:fig1}
\end{figure}

The typical ML workflow consists of three phases (refer to Figure \ref{fig:fig1}): data preprocessing, model building, and model serving \cite{ElMestari2024}.
\begin{itemize}
\item During data preprocessing, raw data is collected, cleaned, and transformed to ensure quality and usability.
\item Model building involves selecting suitable algorithms, training the model, and tuning it to maximize predictive performance. This phase includes iterative testing and validation.
\item Model serving entails deploying the trained model to production, where it generates predictions based on new input data. This stage must consider scalability, efficiency, and ongoing maintenance.
\end{itemize}

\section{Privacy threats}
ML models face a variety of privacy threats, which can be characterized by the assumptions, goals, and capabilities of the adversary \cite{DeCristofaro2020}. Adversaries may be external or internal to the system, with varying levels of access privileges. An internal adversary may exploit insider knowledge, while external adversaries rely on indirect methods. A white-box adversary has full access to model internals (e.g., parameters, architecture), whereas a black-box adversary interacts with the model through an API or interface without insight into its internal workings. Adversaries may engage in passive attacks, such as privacy leakage or active attacks, which alter the system’s operation \cite{Yan2024}. Furthermore, adversaries may follow an honest-but-curious behaviour (complying with system processes but attempting to infer private data), or adopt malicious behaviour, actively seeking to disrupt or manipulate the system.

Privacy threats can target any of the three ML phases: data preprocessing, model training, or inference. While the adversary's primary goal is often to extract users' personal data, attacks may also aim to compromise system integrity through methods such as poisoning or evasion \cite{UstundagSoykan2022}.

The following sub-sections delve into the privacy threats faced by an ML system.

\subsection{Privacy leakage}
ML models are generally optimised by implementing users’ preferences into the training data that may therefore contain private information. Moreover, the inference phase takes user’s inputs unprotected. User’s input queries may contain sensitive or personally identifiable information (PII), or reveal sensitive information about the user when combined with other contextual factors \cite{Mireshghallah2020}. Even seemingly innocuous queries and interactions could indirectly infer user’s personal interests, preferences, behaviour, or characteristics, enabling practices of online tracking, profiling and targeting, including personalised advertisements and recommendations, which raises concerns regarding the invasion of user’s privacy. In several other cases, the underlining ML model itself and its parameters (e.g., architecture, weights, biases, etc.) may require intellectual property protection.

\subsection{Privacy active attacks}
Privacy active attacks imply that the adversary needs to manipulate the ML system by altering the model’s parameters, communications or other system properties to achieve its adversarial goals. For outsourced ML systems e.g., federated learning, distributed learning, or split learning, the adversary may attack the sharing of intermediate computation results. An ML system is vulnerable to several active attacks described in the following.

\subsubsection{Data poisoning attacks}
In data poisoning attacks, called also backdoor attacks, the adversary injects adversarial examples, i.e. poison, into the training set in order to manipulate the behaviour of the model during the training and inference phases. The adversary can also inject poison into the model parameters. Poisoning attacks are not restricted to training data, but also to model weights.

\subsubsection{Membership inference attacks}
If the private information (e.g., training data, inference queries or model parameters) cannot be attained directly, this information can be exposed indirectly using membership inference attacks where the adversary tries to find whether or not a given data instance has contributed to the training of the target model. The success rates of these attacks are due mainly to over-fitted models, but also to the characteristics of the training data like output dimensionality and uniformity within each class. Models tend to memorise rare or unique classes, which is called unintended memorisation, exposing minorities as easy targets for such attacks \cite{ElMestari2024}.

\subsubsection{Property inference attacks}
As a subset of membership inference attacks, property inference attacks aim to infer data patterns or general property of the used training data, e.g., the ratio of examples in each class.

\subsubsection{Attribute inference attacks}
As another subset of membership inference attacks, attribute inference attacks aim to extract sensitive attributes or characteristics of individuals from the target model e.g., demographic information, such as age, gender, or ethnicity.

\subsubsection{Model inversion attacks}
In model inversion attacks, also called data reconstruction attacks, the adversary attempts to reconstruct or reverse-engineer data samples from the training data using the target model or intermediate representations (e.g., for outsourced model). These attacks are facilitated by the transferability property of many ML models and the memorisation capability of deep neural networks (DNNs) \cite{UstundagSoykan2022}. For instance, the adversary can collect query/response pairs and try to recover the model’s parameters. The adversary can also produce random dummy data samples that are fed to the target model to obtain gradients. The dummy data samples are optimised such that the resulting gradients approach the model gradients, and then the dummy data samples will look like the original training data. Another similar technique relies on Generative Adversarial Neural Networks (GANs) to attack the target model \cite{ElMestari2024}.

\subsubsection{Model stealing attacks}
In model stealing attacks, the adversary attempts to steal the target model and extract information about the model’s parameters or internal representations by querying the model and observing its responses.

\subsection{Privacy attacks against large language models}
It has been shown that Large Language Models (LLMs) memorise information in correlation with the number of occurrences in the training data e.g., familiar phrases, public knowledge, templated texts, or other repeated data. These models can also bluntly memorise verbatim sequences of the training data. Data memorisation is called unintended memorisation when the memorised sequences are not directly related to the intended task and are not helpful in improving model accuracy. Another type of memorisation is called counterfactual memorisation which characterises how a model's predictions change if a particular document is omitted during training. 
Because of their memorisation capabilities, LLMs are particularly vulnerable to higher level of data leakage attacks that may reveal sensitive training data. These models suffer also from hallucinations as they may output factually incorrect content.

\section{Countermeasures}
A myriad of privacy-enhancing technologies (PETs) has been proposed to address the aforementioned privacy threats. These technologies can be broadly categorized into three groups: pre-hoc, in-learning, and post-hoc approaches. Pre-hoc and post-hoc methods involve tools applied before or after the training phase and are typically model-agnostic. In contrast, in-learning approaches are integrated directly into the model’s structure and training or inference process.

\begin{figure}[ht!]
	\centering
	\includegraphics[width=0.5\columnwidth]{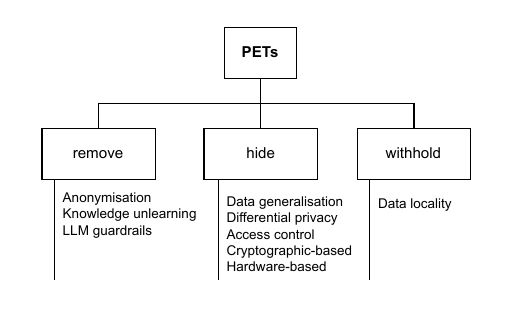}
	\caption{PETs classified based on their privacy-preserving strategy: remove, hide, or withhold private information.}
	\label{fig:fig2}
\end{figure}

Another useful classification of PETs—based on their privacy-preserving strategy—is illustrated in Figure \ref{fig:fig2}. These strategies include removing, hiding, or withholding private information. This latter classification is adopted in the following sub-sections.

\subsection{Remove}
The “remove” strategy involves completely deleting private information from the ML system. Once removed, the data is no longer accessible, ensuring that it cannot be retrieved or misused. The following reviews PETs that follow the removal strategy.

\subsubsection{Anonymisation}
In ML systems, both training and inference data may contain personally identifiable information (PII) or sensitive indicators. Even in the absence of explicit PII, auxiliary information can potentially lead to privacy breaches.

\textbf{Data sanitisation.} A naïve technique to protect privacy consists of removing or generalising PII, such as names, addresses, postcodes, etc., from the training data and inference queries during data preparation. This technique focuses on sanitising data by trying to remove sensitive information while maintaining the statistical trends. However, this technique is susceptible to linkage attacks between different datasets.

\textbf{Pseudonymisation.} Sensitive information can be masked by substituting it with synthetically generated data, non-sensitive placeholders or pseudonyms, which involves removing potentially identifiable information from the data to reduce the risk of data subject identification, while still preserving the structure and relations within the dataset. Moreover, personal identifiers can be replaced with unique pseudonyms, while maintaining the capability for re-identification using lookup tables under controlled circumstances, typically governed by stringent access protocols. Reversible pseudonymisation offers a balanced approach between data utility and confidentiality. However, it may pose technical challenges about protecting the access to lookup tables \cite{ElMestari2024}. There is also some residual risk of subject re-identification, which makes pseudonymisation a weaker form of de-identification compared to anonymisation. Additionally, pseudonyms can be considered personal data in many jurisdictions.

\textbf{Surrogate dataset.} Another anonymisation technique can be achieved by using a surrogate dataset, which is formed by abstracting the dataset using sketching techniques, or more generally generative models. However, generative models require sufficiently large volume of data of good quality and diversity to produce synthetic data. To overcome this challenge, techniques like data augmentation can be used. Data augmentation aims to create duplicates of already-existing data that are modified in order to bring more volume and variety to a given dataset, by using techniques, such as flipping, scaling, cropping, contrasting, adding noise, etc. \cite{ElMestari2024}.

\subsubsection{Knowledge unlearning}
Knowledge unlearning refers to the intentional process of dismissing outdated or incorrect information, mostly concerning LLMs and the ethical implications surrounding users' rights, particularly, the right to be forgotten. The concept of the right to be forgotten gained prominence through the GDPR in Europe, giving individuals the right to delete their data from systems that store and process their personal information. In the context of ML, knowledge unlearning, also known as machine unlearning, aims to selectively forget or remove sensitive information from the model. A straightforward approach to removing a certain information from a model consists of removing this information from the training dataset, and retraining the model, which is not practical for LLMs as their training may take months. Removing this information directly from the model itself (i.e. selectively replacing or deleting model’s weights) is difficult because the model’s parameters are a complex mixture of the whole training dataset [\cite{Zhang2023}. Another approach that provides a more relaxed requirement on targeted information removal, seeks to make the model unable to retain meaningful oversight about the sensitive information, for instance by relabelling the sensitive information in the training dataset with randomly selected incorrect labels or inserting a number of copies of the sensitive information with randomly selected incorrect labels, and then retraining the model on this new incorrect information for some iterations. Existing knowledge unlearning solutions can be categorised into exact machine unlearning that removes sensitive information influence thoroughly from the model and approximate machine unlearning that efficiently minimises sensitive information influence in he model \cite{Xu2023}.

\textbf{Exact machine unlearning.} Exact machine unlearning aims to eliminate the model’s memory of certain information without full model retraining, which is generally achieved using dataset partitioning. 
The Sharding, Isolation, Slicing, and Aggregation (SISA) framework \cite{Bourtoule2019} is a general approach for exact unlearning. SISA consists of an aggregate model of  multiple sub-models trained on disjoint partitions of the training dataset. The removal of a certain information requires only the retraining of the affected sub-models, which are sub-models trained on partitions that contain the sensitive information. SISA includes four key steps \cite{Xu2023}:

\begin{enumerate}
	\item Sharding: The training dataset is divided into multiple separate subsets called "shards" that are used to train sub-models.
	\item Isolation: Sub-models are trained independently on the shards.
	\item Slicing: The sub-models are incrementally trained on partitions of shards, called "slices", in order to track the influence of unlearned data points at a more fine-grained level.
	\item Aggregation: The trained sub-models are aggregated to form the final model, using aggregation strategies like majority voting.
\end{enumerate}

With SISA, the unlearning of a specific data point requires only the retraining of the sub-models associated with shards containing that data point, and starting from the last parameter saved during the slicing step that does not include the data point to be unlearned.

Amnesic unlearning \cite{Graves2020} refers to the process through which ML models can discard specific data points without the need for complete retraining. This is achieved by selectively reversing the learning steps that involved the sensitive data. For this, a list of model's parameter updates from learning on data batches are saved, along with a track record of which examples appeared in which batches. Then, data removal requires only reversing the parameter updates associated wit the batches containing the examples of the sensitive data. A drawback of this approach is the requirement for a large storage space to save the parameter updates. 

\textbf{Approximate machine unlearning.} Approximate machine unlearning seeks to provide a less resource-intensive alternative than exact machine unlearning, by approximating the effect of removing certain data points to an acceptable level. This is achieved by calculating the influence function (a tool of explainable AI e.g., the Newton step method) of data on weights i.e. how the changes in the weights of certain data points affect the model’s parameters. Another approach consists of saving certain information, such as weight updates, during the training for later data deletion. Approximate unlearning leads to a trade-off between machine unlearning completeness and efficiency \cite{Xu2023}. Machine unlearning full completeness is obtained through exact unlearning, whereas better computational efficiency and less storage cost, in addition to higher flexibility, are provided by approximate unlearning. However, challenges remain, such as ensuring the model's accuracy post-unlearning and the risk of potential overfitting if not managed properly.

\subsubsection{LLM guardrails}
Memorisation in the context of LLMs refers to the model's capability to retain specific pieces of information from its training dataset and reproduce them during inference. Unlike traditional learning methods, where generalisation of knowledge is emphasised, LLMs can sometimes recall exact phrases or patterns observed in the training data. To minimise the extent of extractable memorised sensitive content in an LLM model, post-hoc black-box strategies \cite{Zhang2023}, notably LLM guardrails, have been implemented in interactions between the trained model and users, with the objective to monitor and filter the inputs and outputs of the model. LLM guardrails function as a set of programmable, rule-based systems \cite{Dong2024} employed at the query stage, acting as intermediaries between users and LLM models. By implementing a set of predefined guidelines and constraints, these systems help to filter and modify user queries before they are processed by the underlying LLM, thereby enhancing the integrity of the interactions. However, LLM systems may continue to face threats such as prompt injection, prompt extraction, or jailbreak attempts.

Traditional approaches to guardrails often rely on rule-based systems or isolated ML techniques that fall short in addressing the complexity of human language and reasoning. A neural-symbolic design, which integrates both neural agents and symbolic agents, presents a compelling solution. Neural agents, powered by deep learning techniques, excel at recognising patterns in data and generating coherent language outputs. However, they may struggle with abstract reasoning, consistency, and adherence to ethical norms, necessitating the inclusion of symbolic agents. Symbolic agents utilise formal representations of knowledge and logical reasoning to guide decision-making processes. A neural-symbolic approach enables collaborative processing of inputs and outputs to enhance the robustness, interpretability, and ethical alignment of LLMs. Existing guardrail solutions, such as Llama Guard, Nvidia NeMo, and Guardrails AI, use the simplest approach based on loosely coupled neural-symbolic agents.

\subsection{Hide} 
Hiding information means concealing it from view or access, but not necessarily erasing it. Hiding is often used to limit immediate access while retaining the ability to retrieve the information if needed. Different PETs are presented in the following, each achieving the purpose of hiding private information.

\subsubsection{Data generalisation}
Data generalisation is achieved by modifying a dataset so that particular values become more common and groups within the dataset are easier to form.

\textbf{Data aggregation.} To reduce the risk of re-identification, data can be aggregated at a higher level. Aggregation does not aim to alter the information per se, but enables the process of collecting and combining data from multiple sources. However, it may lead to information loss. Moreover, as ML algorithms are becoming more sophisticated, seemingly innocuous data points can be aggregated in a way that may lead to the re-identification of sensitive information.

\textbf{$k$-anonymisation.} Sensitive attributes concealing mechanisms such as $k$-anonymisation technique ensure that personal attributes about a subject are indistinguishable from at least $k-1$ other subjects. The $k$-anonymisation technique has been shown to perform poorly in high-dimensional datasets, which has led to other variants such as $l$-diversity and $t$-closeness. The $l$-diversity technique extends the $k$-anonymisation technique to ensure additional levels of diversity across sensitive datasets, while the $t$-closeness technique improves the $l$-diversity concept \cite{ElMestari2024}. These techniques require to further modify the original data resulting in utility loss. Recently, more research effort focused on utility-preserving anonymisation strategies to overcome the utility loss drawback. 

\subsubsection{Differential privacy}
Differential privacy (DP) aims at revealing statistical information about datasets while protecting the privacy of individual subjects, i.e. the presence or absence of one single subject should not affect the results of a query. To this end, carefully calibrated noise is injected into the statistical computations such that the statistics of the data are maintained while limiting the information inferred about the individual subjects. An algorithm that provides differential privacy is defined as follows: for $\epsilon \ge 0$, an algorithm A satisfies $\epsilon$-differential privacy if and only if for any pair of datasets $D$ and $D$' that differ in only one element, the following equation holds: 
\begin{equation}
	P[A(D) = t] \le e^\epsilon P[A(D') = t], \forall t
\end{equation}
where, $P[A(D) = t]$ denotes the probability that the algorithm $A$ outputs $t$, and $\epsilon$ is a non-negative predetermined value known as the privacy budget. The lower the $\epsilon$, the higher the level of privacy protection, but this may reduce data utility. Adaptive noise mechanisms aim to dynamically adjust noise levels based on privacy budgets and data sensitivity. For example, an adaptive noise mechanism can inject noise into features based on the contribution of each feature to the output, which enables additionally to address fairness issues towards minorities in the dataset and also avoid poor accuracy \cite{ElMestari2024}. For an ML model, the noise can be embedded into the input, the loss/objective function, the gradient updates, the output, and the labels. Generally, DP is mainly used during the training phase. Employing DP during inference is trickier, since it may degrade the accuracy of the model. 

\subsubsection{Ensemble learning}
Ensemble learning is an ML paradigm that aims to improve the performance of predictive models by combining multiple individual models that form the ensemble. These individual models can be of the same type (homogeneous ensembles) or different types (heterogeneous ensembles). The predictions of the individual models can be aggregated using several methods e.g., majority voting, averaging, or stacking (i.e., training a new model on the predictions of the individual models).

\textbf{Private Aggregation of Teacher Ensembles.} The Private Aggregation of Teacher Ensembles (PATE) framework \cite{Papernot2018a} addresses data privacy concerns by harnessing the predictive capabilities of ensemble learning. PATE employs a dual-model structure consisting of multiple "teacher" models and a "student" model. The teachers are trained on disjoint subsets of the data, thereby ensuring that no single teacher has access to all information. This partitioning helps to maintain the confidentiality of the underlying data. Only these teachers have a direct access to data, and these teacher models will not be released into production, instead, they are used to train the student model. The student asks questions to  the teachers, so that the student model is only indirectly trained on data. 

The operation of the PATE framework consists of the following mechanisms:
\begin{itemize}
	\item A set of teacher models is independently trained on different subsets of the training data. Each teacher learns from a distinct partition, which minimises the risk of over-fitting to any specific data point and enhances the diversity of the learned representations.
	\item During the inference phase, the student model receives aggregated predictions from the teacher models rather than direct outputs based on raw data. Specifically, the teachers output their predictions, which are then aggregated, often through a voting mechanism, to provide a final prediction for the student.
	\item To further enhance privacy, the PATE framework incorporates differential privacy techniques. Noise is added to the aggregated predictions from the teachers before they are passed to the student. This noise ensures that the contribution of any single data point remains obscured, thereby providing privacy guarantees that align with differential privacy standards. 
\end{itemize}

\subsubsection{Access control}
Implementing stringent access controls enables to restrict who can view or interact with the private data or the ML model. The access control may follow the zero trust model that requires strict verification of identity and continuously monitors user behavior to detect any suspicious activity.

\textbf{Data privacy vault.} Private ML seeks to develop models that learn from data without compromising the confidentiality of sensitive information. This is achieved through isolation techniques based on access control mechanisms that ensure that data remains protected while still allowing for effective model training and inference. In this respect, an increasingly employed approach involves the use of a data privacy vault \cite{Williams2022} that isolates and protects data from unauthorized access, breaches, and other security threats. In the context of ML, a data privacy vault may feature techniques such as data anonymisation and pseudonymisation to ensure obfuscated PII. Additionally, it may include access control mechanisms using a combination of a zero trust architecture and role-based or attribute-based access control mechanisms that facilitate secure data sharing without exposing sensitive information.

\subsubsection{Cryptographic-based approaches}
The integration of cryptographic tools into ML processes can significantly bolster privacy, by offering a strong confidentiality guarantee to ML applications, also known as "confidential-level privacy" \cite{ElMestari2024}. In terms of performance, cryptographic techniques often suffer from computational inefficiency. Since, the inference process is computationally less consuming than the training process, the more advanced cryptosystems (e.g., homomorphic encryption) are applied during the inference phase than for training. Standard classical cryptographic algorithms (e.g., RSA, AES, etc.), in addition to post-quantum cryptographic algorithms, can be employed to provide confidentiality and integrity protection for data in transit and at rest. The following focuses on more advanced cryptographic primitives used to enhance privacy in ML applications. 

\textbf{Secure multi-party computation. }Secure multi-party computation (MPC) enables multiple parties (often mutually distrusting) to collaboratively compute a function over their inputs without revealing their respective inputs. There are different techniques to achieve secure MPC, including arithmetic secret sharing, Boolean secret sharing, Yao’s garbled circuits, and homomorphic encryption. Zero-knowledge proofs can be used by MPC in some cases for verification of task output. In the context of ML, secure MPC allows different parties to train a model on combined datasets without revealing their proprietary data to one another. For instance, in federated learning, secure MPC enables secure aggregation of local models, allowing parties to conjointly train a shared model. During model inference, secure MPC enables to ensure privacy by performing computations on encrypted data, protecting sensitive information from central servers. Additionally, secure MPC can facilitate secure data labelling by enabling multiple parties to label data collaboratively without exposing raw labels. 
The field of secure multi-party computation is more mature than other encryption techniques like homomorphic encryption.

\textbf{Private set intersection.} Private Set Intersection (PSI)  \cite{Lu2020} is a cryptographic protocol that enables two or more parties to compute the intersection of their data sets without revealing any other information about their individual data. PSI allows multiple parties to collaborate on data analysis, for instance de detect and mitigate biases, without exposing sensitive information. In federated learning, PSI can be utilised to ensure that only the necessary data required for model updates is shared, enhancing both efficiency and security.

\textbf{Homomorphic encryption.} Homomorphic encryption (HE) enables to perform computation over encrypted data.      HE schemes consist of multiple types of constructions that enable different types of computations represented as either Boolean or arithmetic circuits, over encrypted data. These schemes can be divided into the following common categories:
\begin{itemize}
	\item Partially Homomorphic Encryption (PHE): PHE supports the evaluation of circuits consisting of only one type of gate, e.g., addition or multiplication.
	\item Somewhat Homomorphic Encryption (SWHE): SWHE supports the evaluation of two types of gates, but only for a subset of circuits.
	\item Levelled fully homomorphic encryption (LHE): LHE supports the evaluation of arbitrary circuits composed of multiple types of gates of bounded (pre-determined) depth.
	\item Fully Homomorphic Encryption (FHE): FHE supports the evaluation of arbitrary circuits composed of multiple types of gates of unbounded depth. FHE is considered as the strongest notion of HE.
\end{itemize}
Examples of HE schemes include Paillier’s encryption and exponential ElGamal encryption schemes. The first plausible construction of an FHE scheme was proposed by Craig Gentry, using lattice-based cryptography, in 2009.

In the context of ML, HE \cite{PulidoGaytan2021} have some limitations in terms of computation functions. For instance, training some ML algorithms may require computations in floating-point, which can be mitigated by scaling the floating-point numbers up to integers. Moreover, training a deep neural network model requires performing a large chain of multiplications, in addition to non-linear function evaluations that cannot be implemented by HE. Techniques that address these issues include the use of lookup tables or low-degree polynomials as substitutes. Lookup tables suffer from the problem of exploding noise budgets, which can partially be tackled using bootstrapping techniques \cite{ElMestari2024}. On the other hand, techniques employing low-degree polynomials enable to concert activation functions (e.g., ReLU, Sigmoid, and Tanh) in neural networks \cite{Mireshghallah2020} and operators such as SoftMax and LayerNorm in the transformer model \cite{ElMestari2024} into their polynomial equivalent. The coefficients of these polynomials can be learnt during model training (e.g., using knowledge distillation). The ciphertext of an HE scheme is usually much larger than the plaintext, and the computations on the ciphertext take longer time than those on the plaintext. To alleviate this problem, one technique consists of encoding several data messages into one single plaintext and then using single instruction multiple data (SIMD) method (e.g., based on the Chinese Reminder Theorem), to process these messages in batch with no extra cost \cite{ElMestari2024}. Even though the performance of HE schemes have been improved in the last decade, deploying HE techniques still imposes a significant computational hurdle. Since, model inference requires less computational resources than the training, HE is often applied during the inference phase than during model training.

\textbf{Functional encryption.} Functional encryption (FE) \cite{Panzade2022} is a generalisation of public key encryption, allowing a user to encrypt data such that a specific function can be computed on the encrypted data without revealing the underlying plaintext. FE includes the special cases of attribute-based encryption (ABE) and identity based encryption (IBE). The key difference between FE and HE is that HE produces ciphertext results, whereas FE produces plaintext results. In the context of ML, FE significantly reduces the risk of exposing sensitive information by allowing computations on encrypted data. For this, the server must obtain FE secret keys in order to execute the FE function. Current FE schemes are often limited to specific types of functions, and can be categorised as either inner-product or quadratic FE-based approaches \cite{Panzade2022}. Advancements are needed to broaden the range of FE functions. The overhead associated with FE computations can be significant, which may hinder its practical adoption in some ML applications, especially real-time applications. Moreover, the generation and distribution of FE keys must be managed securely to prevent unauthorised access.

\textbf{Zero-knowledge proofs.} Zero-knowledge proofs (ZKPs) are two-party cryptographic protocols that enable one party (the prover) to prove to another party (the verifier) that a given statement is true without revealing the data necessary to prove the statement or any additional information beyond the validity of the statement itself. ZKP can be implemented using cryptographic primitives, such as bilinear pairings, ore more simpler hash and commitment schemes, and through the utilisation of blockchain and consensus protocols \cite{Xing2023}. In the context of ML, the foundational principles of ZKPs can be applied to ensure that models can be trained and evaluated while maintaining the privacy of the data. For instance, a model can be trained on encrypted data, sharing insights without exposing the original data. Moreover, ZKPs enable to prove that the models have been trained on valid datasets without revealing the datasets themselves, which is particularly useful in outsourced learning scenarios, such as federated learning and cloud-based learning. In these scenarios, ZKPs can enable the verification of global model aggregation integrity while preserving the local models’ privacy. During the inference phase, ZKPs can be a valuable tool for providing private predictions by ensuring that neither the input data nor the model parameters are compromised. Depending on their constructions, ZKPs can introduce a significant computational overhead, making the process slower and more resource-intensive than traditional approaches.

\textbf{Query obfuscation.} Query obfuscation refers to the process of transforming a user's input queries in such a way that the intent or content of the query is obscured while still allowing the underlying ML model to generate relevant outputs. The objective is twofold: to protect sensitive information from unauthorized access and to prevent potential adversaries from inferring patterns or details about users based on their queries. In this domain, prominent protocols include Privacy Information Retrieval (PIR) and Oblivious Transfer (OT).

Privacy Information Retrieval (PIR) protocols \cite{Vithana2023} enable a user to retrieve data from a database without disclosing which specific data item is being queried. Conceptual extensions of PIR include private set intersection (PSI), private set union (PSU), and private read-update-write (PRUW) intended to solve privacy issues associated with distributed learning applications. An information-theoretically secure PIR protocol requires either multiple non-colluding servers holding copies of the database, or an amount of communication at least the size of the database. Existing PIR protocols suffer from computational issues, whereas OT protocols offer a scalable and efficient solution. Additionally, OT adds an additional privacy requirement that the user does not learn any item other than the requested item, which provides a symmetry in terms of privacy between the user and the database. Therefore, more research has been dedicated to OT. 

OT is a type of two-party protocol that facilitates the exchange of information while preserving the privacy of both parties involved. In a standard form, one party, referred to as the sender, possesses multiple pieces of information (e.g., messages), and the other party, known as the receiver, wishes to obtain one of these pieces. The key characteristic of this transfer is that the sender remains unaware of which message was chosen by the receiver, while the receiver learns only the selected message and nothing more. In the context of ML, OT enables oblivious predictions i.e. the model predictions are computed such that the deployment server that computes the prediction based on the client input, does not learn anything about the client input and the client does not learn anything about the used model, except for the prediction result \cite{ElMestari2024}. OT is often more computationally efficient than HE, at the cost of more communications. An oblivious neural network \cite{Liu2017} is a neural network that supports privacy-preserving predictions. In a typical construction, the input and output of each layer of the oblivious network is shared between both the client and the server such that combination of the shares is equal to the input and output to that layer in the non-oblivious version of the network. Hardware-based approaches. Hardware-based approaches play an important role in enhancing the privacy of data processed by ML applications. These approaches can be used as performance accelerators for different resource-consuming PETs like homomorphic encryption. Additionally, they can be used to provide trusted computations.

\subsubsection{Trusted computing}
The integration of trusted computing (TC) technologies into ML frameworks offers promising solutions to mitigate privacy concerns. TC refers to a set of hardware and software technologies, such as Trusted Platform Module and Trusted Execution Environment (e.g., Intel's Software Guard Extensions or ARM's TrustZone), designed to provide a secure environment for computing operations.

The core concept of TC is the chain of trust that is established by validating each hardware and software component, starting from the root of trust, a source secure by design, to hardware platform, operating system, and applications. using typically chained hash functions. Other TC concepts include trusted boot, sealed storage, curtained memory, attestation, integrity measurement and secure I/O \cite{Cybersecurity.2023}. Originally, TC relied on a separate hardware module, the Trusted Platform Module (TPM), which is a dedicated microcontroller designed to protect hardware using integrated cryptographic keys. However, a TPM comes a with a reduced functionality restrained by a predefined set of APIs. It does not provide, for instance, the capability to securely execute third party applications. On the other hand, a Trusted Execution Environment (TEE) offers an isolated execution environment using a hardware isolation mechanism, such as Intel's Software Guard Extensions (SGX) or ARM's TrustZone, that enables the tamper-resistant execution of arbitrary code within a confined environment. 

TEE is an attractive approach to provide the integrity and confidentiality of data in use against malware, but also against super-privileged software and users e.g., the operating system, the hypervisor, the system administrator or the cloud provider. Even though, TC technologies are prone to information leakage attacks, such as side channel attacks on hardware implementation (e.g., power consumption, electromagnetic radiations), these attacks can be generally mitigated using additional oblivious techniques \cite{ElMestari2024}. Since TC technologies enable to achieve both data protection and secure computation, they provide an opportunity for ML to protect sensitive data in use and at rest, and secure learning tasks. The whole or parts of ML tasks (training or inference) can be loaded in a trusted environment to be attested and verified.

\subsection{Withhold}
Withholding private information refers to the deliberate decision of not sharing certain data. A noteworthy approach employed to withhold data is through data locality that refers to the principle of processing data close to where it is generated.

\subsubsection{Data locality}
By keeping sensitive data within its originating environment (i.e. at user's device), data locality minimises the need for direct data transmission and sharing across networks, thereby reducing exposure to potential security breaches during transit. This approach enhances control over data, reinforcing data protection and privacy. Two main techniques based on data locality include federated learning and split learning. 

\textbf{Federated learning.} Federated learning (FL) or collaborative learning is a decentralised ML technique that facilitates the general approach of "bringing code to data instead of bringing data to the code" \cite{Farahani2023}. FL enables multiple parties, called clients, to collaboratively train a model while ensuring that their data remains private. A central server is generally used to orchestrate and coordinate the learning process. The clients train models locally using their own data. Instead of sending raw data to the server, clients compute updates to their models based on local datasets and share only those updates (model weights or gradients) with the server. The server aggregates the updates to improve a global model without ever accessing the individual datasets. This distributed approach significantly enhances privacy, as personal data remain on individual devices, allowing compliance with privacy regulations and user consent policies. However, malicious servers may, in some cases, extract sensitive information from shared local models (e.g., gradients). To mitigate this privacy breach, FL often integrates additional PETs (e.g., differential privacy, homomorphic encryption).

\textbf{Split learning.} Split learning (SL) provides a complementary strategy to FL by enabling collaborative learning between clients and a central server while preserving data privacy. In this approach, a neural network is partitioned into two segments: one part runs on a client device and the other on other clients or a server. The client processes its local data through the initial layers of the neural network up to the cut layer. The output of the cut layer, referred to as smashed data, is sent to another client or a server that completes the training. Back-propagation is realised in the same fashion. To maintain data locality, the intermediary clients or the server do not have access to the raw data, only the smashed data, during the forward or back propagation. This method allows clients to retain full control over their raw data, as it is never exposed to remote servers. Moreover, SL reduces communication overhead since only smaller outputs are transmitted, making it both a privacy-preserving and bandwidth-efficient solution.

\textbf{Off-site Tuning.} Off-site Tuning \cite{Xiao2023} is an approach that aims at tuning a large model to a specific task using user's data., while preserving the privacy of both the user's data and the model. In Off-site Tuning, a lightweight adapter fine-tuned locally on the user's data is transmitted to the model's owner, thus ensuring that sensitive data remains under the user's control. On the other hand, the model's owner sends only a compressed emulator of the model to the user for the tuning. 

The different discussed PETs are compared in table \ref{tab:table}, highlighting the delicate trade-off between guaranteeing privacy and minimising utility loss and cost of applying PETs into ML systems.

\begin{table*}[ht!]
  \centering
	\caption{Qualitative comparison between PETs}
	\begin{tabular}{ clcccc } 

\hline
		\textbf{PET class} & \textbf{PET technique} &	\textbf{Training(T)\slash Inference(I)} & \textbf{Privacy guarantees} & \textbf{Utility loss} & \textbf{Cost} \\

\hline
		\multirow{3}{4em}{\textbf{Remove}} & Anonymisation & T & Low & Medium & Low \\
		& Knowledge unlearning & T & High & High & Medium \\
		& LLM guardrails & I & Medium & - & Low \\
\hline
		\multirow{5}{4em}{\textbf{Hide}} & Data generalisation & T & Low & High & Low \\
		& Differential privacy & T\slash I & High & High & Low \\
		& Access control & T\slash I & Medium & - & Low \\
		& Cryptographic-based & I & High & Low & High \\
		& Hardware-based & T\slash I & High & Low & High \\
\hline
		\textbf{Withhold} & Data locality & T & Medium & Medium & Low \\
\hline
	\end{tabular}
	\label{tab:table}
\end{table*}

\section{Challenges}
One of the central promises of privacy-enhancing technologies (PETs) in AI is the ability to balance the utility of AI systems with the imperative of safeguarding individual privacy. However, several critical challenges must be addressed to fully realise this potential. The following outlines key obstacles in the integration of PETs into AI workflows:

\begin{itemize}
    \item \textbf{Algorithmic complexity.} The integration of PETs often necessitates substantial modifications to existing AI algorithms. For example, techniques like differential privacy require careful calibration of privacy budgets to maintain a trade-off between model performance and privacy guarantees.
    
    \item \textbf{Computational overhead.} Many PETs introduce significant computational burdens that can hinder scalability. For instance, homomorphic encryption enables operations on encrypted data but incurs substantial latency compared to plaintext computation. This performance cost is especially problematic in real-time applications or resource-constrained environments. Research into lightweight cryptographic schemes and adaptive privacy-preserving mechanisms is ongoing to mitigate this issue.
    
    \item \textbf{Utility degradation.} Enhancing privacy often comes at the cost of reduced model accuracy. Perturbation techniques, such as noise addition, can degrade data fidelity to the extent that trained models underperform relative to those trained on unmodified data. Finding an optimal balance between privacy protection and model utility remains a persistent challenge.
    
    \item \textbf{Limited collaboration.} In collaborative learning scenarios (e.g., federated learning), PETs restrict the sharing of raw data, limiting the scope of cross-organisational or cross-device learning. While these technologies preserve confidentiality, they can also constrain the richness and diversity of the data accessible for model improvement.
    
    \item \textbf{Evolving threat landscape.} As PETs advance, so do adversarial techniques aimed at circumventing them. Attacks such as membership inference or model inversion pose serious threats to privacy guarantees, underscoring the need for continual evaluation and enhancement of PET robustness.
    
    \item \textbf{Lack of standardisation.} The field currently lacks universally accepted standards and best practices. As a result, PET implementations vary in efficacy and interoperability across different platforms. Developing standardised frameworks that support scalability, compatibility, and verifiability remains an open research challenge.
\end{itemize}

\section{Conclusion}
In the era of data-driven AI, ensuring the privacy of individuals while maximising the utility of models has become an essential objective. Privacy-enhancing technologies (PETs) offer a promising pathway toward achieving this balance, enabling data to be leveraged in a secure, ethical, and compliant manner.

Technologies such as differential privacy, knowledge unlearning, cryptographic methods, hardware-based isolation, and data locality represent leading approaches in the quest for privacy-preserving AI. These techniques help align AI development with the principle of \textit{privacy by design}, fostering user trust and regulatory compliance.

While many PETs remain in early stages of maturity and face significant practical challenges—ranging from computational overhead to implementation complexity—their continued evolution is critical. Ongoing research and development efforts will be essential to improve scalability, standardisation, and usability, ultimately enabling widespread adoption.

As AI continues to permeate diverse domains, embedding privacy as a core design principle will be instrumental in promoting responsible innovation and sustained societal trust in intelligent systems.

\bibliographystyle{unsrt}  
\bibliography{references}  

\end{document}